# Evidence for a re-entrant character of magnetism of σ-phase Fe-Mo alloys: non-linear susceptibilities

S. M. Dubiel[*]

AGH University of Science and Technology, Faculty of Physics and Applied Computer Science, al. A. Mickiewicza 30, PL-30-059 Kraków, Poland

## Abstract

Non-linear ac magnetic susceptibility terms viz. quadratic $\chi_2$ and cubic $\chi_3$ were measured versus temperature and frequency for a series of the σ-phase $Fe_{100-x}Mo_x$ (47≤x≤53) compounds. Clear evidence was found that the ground magnetic state of the samples is mixed i.e. constituted by two phases: a spin glass (SG) and a ferromagnet (FM), hence the magnetism of the investigated samples can be regarded as re-entrant. Based on the present data, previously reported magnetic phase diagram has been upgraded [J. Przewoznik, S. M. Dubiel, J. Alloy. Comp., 630 (2015) 222].

[*]Corresponding author: Stanislaw.Dubiel@fis.agh.edu.pl

## 1. Introduction

The sigma ($\sigma$) phase (space group $D^{14}_{4h}$ - $P4_2/mnm$ ) represents the Frank-Kasper (FK) family of phases known also as topologically close-packed structures [1]. So far it has been discovered in alloys in which at least one constituting element is a transition metal. A unit cell of $\sigma$ is tetragonal and it accommodates 30 atoms distributed over 5 different lattice sites. Their coordination numbers are high (12-16), and their occupancy is not stoichiometric. These features, jointly with the fact that $\sigma$ can be formed in a certain composition range, make the $\sigma$-phase alloys highly disordered. The latter causes a wide diversity of physical properties that can be tailored by changing constituting elements and/or their relative concentration. The interest in $\sigma$ is additionally augmented by its detrimental effect on useful properties of industrially important materials (steels) in which it has precipitated e. g. [2,3]. On the other hand, attempts have been undertaken to profit of its high hardness in order to strengthen the materials e. g. [4,5].

Magnetic properties of $\sigma$ in binary alloys, the subject of the present paper, were so far revealed in Fe-Cr, Fe-V, Fe-Re and Fe-Mo systems [6-12]. Its magnetism has been recently shown to be more complex than initially anticipated viz. in all four cases the ground state is constituted by a spin glass (SG) which in the case of the Fe-Cr and Fe-V systems has a re-entrant character [11]. Some magnetic features revealed for $\sigma$ in the Fe-Re [9] and Fe-Mo systems [11,12] indicate that it also may be the case here. It is known that non-linear AC susceptibilities, $\chi_2$, $\chi_3$,..., can give a valuable information on the character of magnetism. The magnetization $M$ in a weak magnetic field $H$ can be expended in powers of the field:

$$M = M_o + \chi_1 H + \chi_2 H^2 + \chi_3 H^3 + \ldots\ldots \qquad (1)$$

Where $M_o$ is the spontaneous magnetization. In a ferromagnet (*FM*) all terms are expected to have a non-zero contribution to $M$, while in canonical SGs ($M_o$=0) only odd-terms should be present [13,14]. Consequently, a temperature behavior of the non-linear susceptibilities can be used not only to uniquely characterize *FMs* e. g. [15-17], but also to make a distinction between the canonical, cluster and re-entrant *SGs* e. g. [18]. For the latter two phases viz. one with the *FM*-order and the other one with the *SG*-order co-exist [19]. Correspondingly, the non-linear susceptibilities are expected to exhibit features characteristic of both magnetic states. Indeed, observation of $\chi_2$ and $\chi_4$ in La$_{0.7}$Pb$_{0.3}$(Mn$_{1-y}$Fe$_y$)O$_3$ magnetite confirmed such expectation [18]. Systematic ac magnetic susceptibility measurements carried out on a series of $\sigma$-Fe$_{100-x}$Mo$_x$ (*x*=45, 47, 51, 53) samples gave a clear evidence that the ground magnetic state of the $\sigma$-FeMo alloys was constituted by *SG* [12]. Its features, despite a high concentration of the magnetic carriers, were found to be close to those typical of the canonical *SG*s. On the other hand, **dc** magnetization and Mössbauer spectroscopic measurements revealed that a magnetic ordering occurred at temperatures significantly higher than the spin-freezing ones [10]. The latter feature suggests a magnetic order exists above the spin-freezing temperature, hence the magnetism of the system has a re-entrant character. To shed more light on the issue, non-linear susceptibilities were measured on the same samples as reported elsewhere [12], and the results obtained are presented and discussed in this paper.

## 2. Experimental

The ac magnetic susceptibility measurements data were performed using the Quantum Design physical property measurement system (PPMS) between 2.0 and 100 K in a 2 Oe ac magnetic field (and zero external dc magnetic field) for frequencies varying from 10 Hz to 10 kHz. The data were collected during the cooling down and warming up runs.

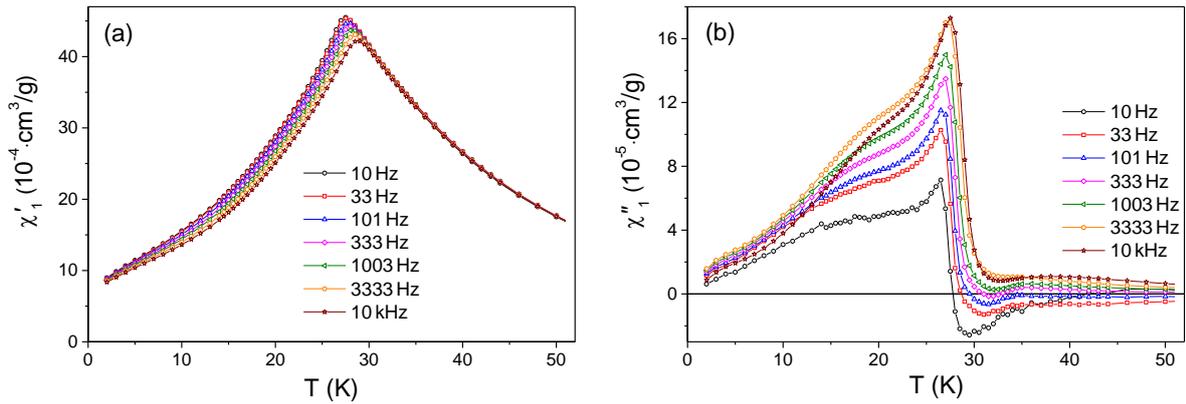

Fig. 1 (color online) (a) Real, $\chi_1'$, and (b) imaginary, $\chi_1''$, parts of ac magnetic susceptibilities recorded on the σ-$Fe_{53}Mo_{47}$ alloy vs. temperature, *T*, and for various frequencies, *f*, shown.

Examples of the linear ac magnetic susceptibility curves, $\chi_1'(T)$, are shown in Fig. 1a. They give a clear-cut evidence that (1) a well-defined cusp, whose position, $T_f$, defines a spin-freezing temperature , and (2) the cusp shifts towards higher temperature with growing frequency, *f*. These features have been regarded as the most important signature of a spin glass (*SG*) state. The $\chi_1''(T)$-curves, despite lower intensities, have a more diversified structure viz. display two characteristic temperatures: a higher one (rather well-defined) which we will call $T_{if}$, and a lower one (ill-defined) which we will call $T_{si}$ and will associate it with the upper limit of a strong-irreversibility state of *SG*. In other words, the *SG* state in the studied samples seems to be magnetically heterogeneous.

## 3. Results and discussion

### 3.1. Tholence criterion

The frequency dependence of the spin-freezing temperature, $T_f$, was reported and discussed in detail in our previous paper [12]. Values of characteristic quantities were found to be close to those typical of the canonical *SG*s. For example a relative shift of $T_f$ per a decade of frequency was in the range of 0.0122-0.0135, the activation energy span between 32 and 65 K, the dynamic exponent had values within 8 and 10 range, and the characteristic frequency, $f_o$, obtained from the critical slowing-down law varied between $10^{13}$ and $10^{15}$ Hz. This rather-unexpected behavior (because of a very high concentration of magnetic entities) finds further support in form of the Tholence criterion which has been regarded as a measure of a degree of clustering [20]. The corresponding figure of merit, $\Delta_C$, is defined as follows:

$$\Delta_C = \frac{T_f - T_o^{VF}}{T_f} \qquad (2)$$

Where $T_o^{VF}$ is a Vogel-Fulcher temperature. For the studied samples $\Delta_C$ spreads between 0.04 and 0.105 i.e. the degree of the clustering is similar to that found in the canonical *SG*s [21]. Apparently, the highly itinerant character of magnetism in the σ-phase compounds [9,22], hence very long-range magnetic interactions, causes that the greatly Fe-concentrated alloys, as the presently studied ones, show features representative of well-behaving *SG*s.

### 3.3. Non-linear susceptibility

We start the presentation and discussion of the non-linear susceptibilities with $\chi_3$ (real part). Its temperature dependence for all samples and two chosen frequencies viz. $f$=333 Hz and 3.33 kHz are displayed in Fig. 2. Theoretical calculations predict that $\chi_3$ is expected to be present both in ferromagnets as well as in spin glasses [13-15,17]. However, in the latter case $\chi_3 < 0$ for all temperatures, and it diverges at $T_f$ e. g. [14], while for *FM* $\chi_3 > 0$ in the ferromagnetic state ($T<T_C$) and $\chi_3< 0$ in the *PM* phase ($T> T_C$) with a divergence at the Curie temperature, $T_C$

e. g. [17]. Figure 2 gives a definitive evidence that $\chi_3 < 0$ for all four samples. All curves show minima that can be associated with the spin-freezing temperature which we denote here as $T_{3f}$ and show its values in Table 1. They are similar to the values determined from the linear part of $\chi$, $T_f$, as reported elsewhere [12], although a small enhancement can be observed due to higher frequencies. The latter can be used to distinguish not only between spin glass and super paramagnetic states but also to differentiate between various types of *SG*s. For that purpose the following figure of merit has been used:

$$RST = \frac{\Delta T_f / T_f}{\Delta \log f} \qquad (1)$$

Values of *RST* obtained from $T_{3f}$ are denoted by $RST_{3f}$ and displayed in Table 1. They are by factor 2-5 higher than the corresponding ones obtained using the values of $T_f$ yet characteristic of well-behaving *SG*s. In the light of the high concentration of Fe atoms, it seems rather paradoxical. One has, however, to remember that for a magnetic behavior of a system, and, in particular, for the one with a spin glass ordering, not only a concentration of magnetic carriers but also a range of interaction between them are crucial. If the latter is a long range, like in the present case (due to an itinerant character of magnetism in the σ-phase compounds), then clusters of magnetic atoms which are chemically expected are magnetically ill-defined. Consequently, a chemically concentrated systems, like the ones in this study, do not behave like cluster spin glasses but their behavior rather resembles that of the canonical ones. Regarding an intensity of the $\chi_3$-curves it strongly decreases with the increase of Mo content and the $T_{3f}$-values shift towards lower temperature. These features are in line with our previous observations [12] according to which the strength of magnetism in the investigated samples significantly weakens when the Mo content increases. In any case, the data presented in Fig. 2 agree with the predictions for a spin glass. However, as suggested in Ref. 18, the temperature behavior of the even terms in eq. (1) is better suited for an unambiguous distinction between the ferromagnetic and spin glass orderings. The even terms, and, in particular $\chi_2$ and $\chi_4$ are especially well-suited to make a distinction between pure (canonical) spin glasses and re-entrant ones in which the *FM* and *SG* orderings co-exist [19]. This is especially important in the present case where a cluster or re-entrant SG behavior is expected due to the high concentration of Fe atoms, a (ferro) magnetic ordering occurs at $T > T_f$ whereas various figures of merit are typical of the canonical SGs.

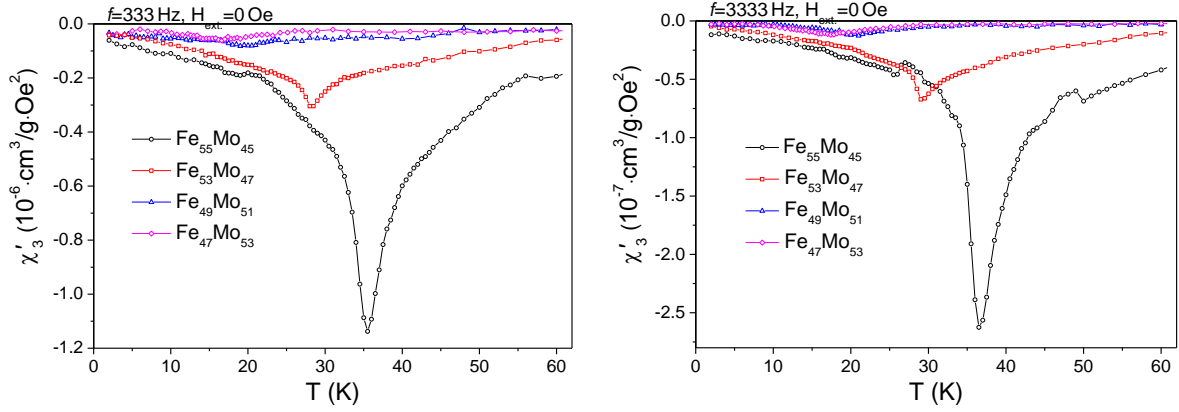

Fig.2 Real part of $\chi_3$ versus temperature as measured for the studied σ-phase samples at two different frequencies shown. Note a strong dependence of the signal on the composition.

Table 1

Characteristic temperatures in K for the investigated samples of σ-$Fe_{100-x}Mo_x$, as explained in the text, and the figure of merit, *RST*, as defined by eq. (1).

| x | $T_f$ [12] | $T_{3f}$ (333) | $T_{3f}$ (3333) | $T_C$ (dc) [10] | $T_C$ (ac) | $RST_f$ [12] | $RST_{3f}$ |
|---|---|---|---|---|---|---|---|
| 45 | 34.5 | 35.5 | 36.7 | 44.5 | 39.3 | 0.013 | 0.034 |
| 47 | 27.5 | 28.55 | 29.3 | 33.0 | 30.7 | 0.0135 | 0.026 |
| 51 | 19.2 | 19.4 | 20.65 | 21.5 | 22.0 | 0.012 | 0.064 |
| 53 | 16.15 | 16.5 | 17.35 | 20.0 | 18.8 | 0.013 | 0.051 |

The two phases have different time scales (relaxations): slow for *SG* and fast for *FM*. The ac susceptibility method offers a suitable tool to investigate such mixed states: at low frequencies the *SG* phase should be visible while at high frequencies the *FM* phase should be detectable. Concerning $\chi_2(T)$, the first even term, it should be absent for the pure *SG*, whereas for the *FM* it should be negative for $T < T_C$ (with a divergence at $T_C$) and equal to zero for T > $T_C$ [17]. Consequently, for a system in which *SG* and *FM* co-exist, one should observe a frequency-induced change of behavior of $\chi_2$ between the two cases.

The data obtained for two samples with the highest content of Mo viz. σ-$Fe_{47}Mo_{53}$ and σ-$Fe_{49}Mo_{51}$ agree perfectly with this expectation (Fig. 3a) i.e. $\chi_2 < 0$ for the two highest frequencies i.e. 3.33 and 10 kHz, while for lower frequencies $\chi=0$ within error limit. In other words, two magnetic orderings viz. *SG* and *FM* co-exist in the ground magnetic state of these two samples. Two other samples viz. with the lowest concentration of Mo i.e. σ-$Fe_{55}Mo_{45}$ and σ-$Fe_{53}Mo_{47}$ show a different, thus unexpected behavior.

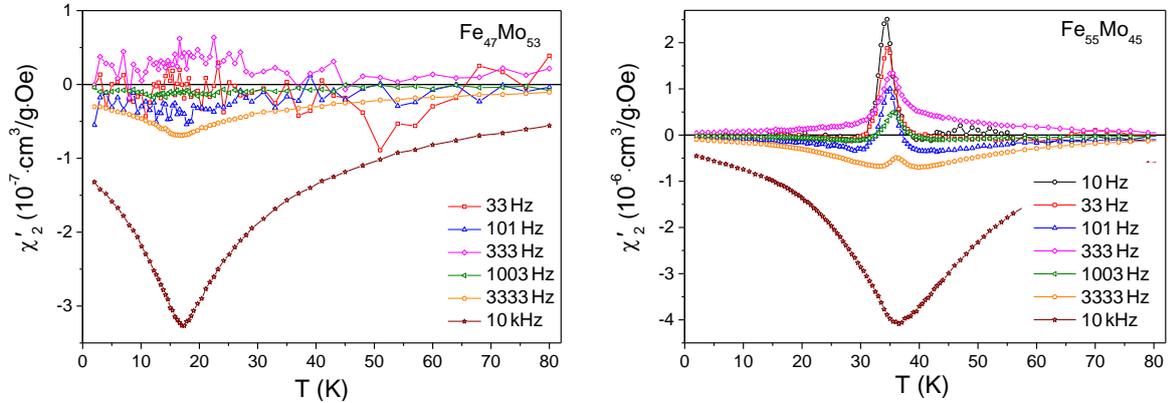

Fig. 3 Temperature dependence of $\chi_2$ for various frequencies as detected for two samples of the σ-phase with the extreme compositions.

While for the highest frequency (10 kHz) $\chi_2$ obeys the mean-field prediction i.e. is negative and has a minimum, for lower frequencies an unexpected change occurs viz. a well-defined maximum instead of the minimum appears, it grows with the decrease of *f* and, eventually, $\chi_2$ becomes positive in the entire range of temperature. This behavior also indicates the heterogeneous or mixed ground state of the two samples, one component having the *FM* ordering. However, a nature of the component represented by the curves with maximum is unclear in the light of existing predictions, and, in particular, those based on the mean-field treatment that does not account for magneto-crystalline and shape anisotropies [17]. Anyway, it should be mentioned that a positive, peak-like shaped $\chi_2$ curve was also observed in a ferromagnetically ordered $Ga_{0.6}Mo_2S_4$ spinel [15]. It is therefore reasonable to assume that in systems with *RSG* ordering $\chi_2$ can be either positive or negative depending on the frequency and sizes of a ferromagnetically ordered component.

### 3.4. Magnetic ordering above $T_f$

The existence of the ferromagnetic sub-phase in the ground state, positive values of the Curie-Weiss constant [12] as well as the Mössbauer spectroscopic and *dc* magnetization measurements at [10] testify to the existence of a ferromagnetic ordering below the paramagnetic and above the *RSG* phases. The temperature of this ordering can be also determined as an inflection point in the linear *ac* susceptibility curves for $T>T_f$. The values of temperature obtained in this way are labelled as $T_C$ (ac) and presented in Table 1 together with the corresponding values found from the dc magnetization curves, $T_C$ (dc) [10]. A reasonably good agreement can be seen, especially for $x$=51 and 53.

### 3.5. Magnetic phase diagram

The presently reported and discussed results permit upgrading the magnetic phase diagram of the σ-FeMo alloy system which was suggested in our recent paper [12]. In particular, considering the non-linear ac susceptibilities demonstrated that the ground state is a mixture of a spin glass and a ferromagnetic phases, hence it can be termed as a re-entrant spin glass. This feature in combination with positive values of the Curie-Weiss temperature, as found in our previous publication [12], permits to conclude that the magnetic state that occurs between the paramagnetic one and the *RSG* phase has a ferromagnetic ordering. The improved version of the phase diagram is presented in Fig. 4.

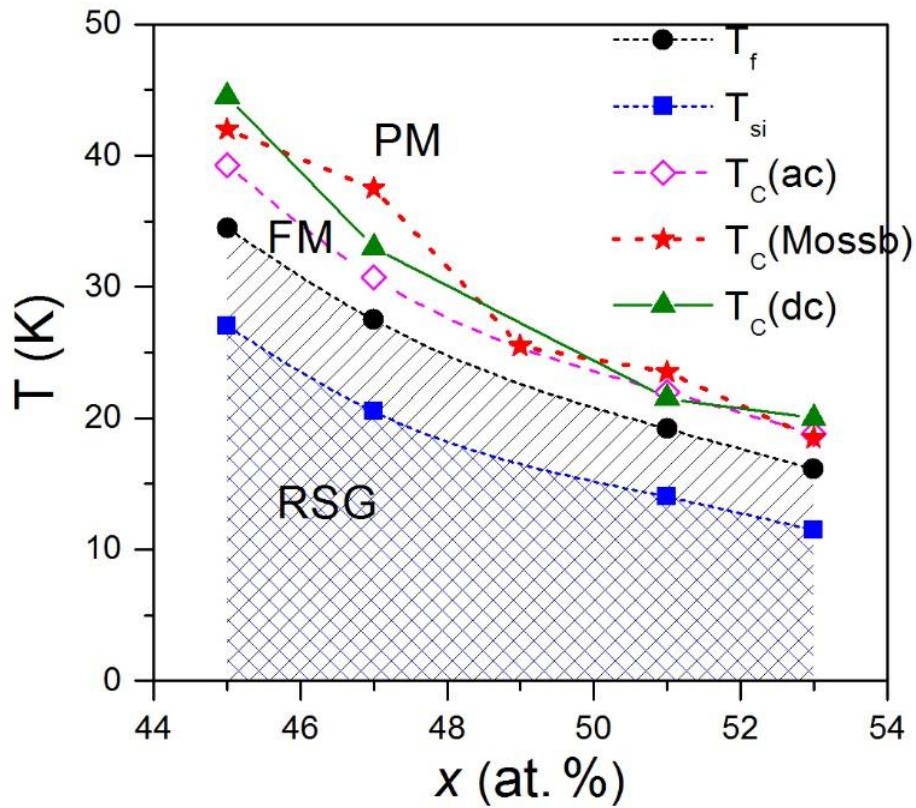

Fig. 4 The upgraded magnetic phase diagram of the σ-$Fe_{100-x}Mo_x$ alloys relative to its version presented in [12]. Added are the $T_C$-values obtained from the infection point of the *ac* susceptibility curves, $T_C(ac)$, as well as those determined from the temperature dependence of the average hyperfine field, $T_C(Moss)$, as reported in [10]. The lines are to guide the eye. The range of the RSG regime is divided into the one with a strong irreversibility (diagonal crossing) and that with a weak irreversibility (diagonal-bottom-top).

## 4. Summary

Non-linear ac magnetic susceptibility terms viz. the quadratic, $\chi_2$, and the cubic, $\chi_3$, have been considered for a series of the σ-$Fe_{100-x}Mo_x$ alloys ($45 \leq x \leq 53$). They clearly indicate that the magnetic ground state is not homogeneous but it is a mixture of two phases having spin glass and ferromagnetic orderings, respectively. In other words, the magnetism of the investigated samples has a re-entrant character. This finding is in line with our previous observations concerning the magnetism in other σ-phase binary alloys viz. Fe-Cr, Fe-V and Fe-Re [9,11]. An upgraded magnetic phase diagram has been outlined for σ the studied alloy system.

**Acknowledgements**

This work was supported by The Ministry of Science and Higher Education of the Polish Government. J. Przewoźnik is acknowledged for performing the susceptibility measurements and fruitful discussions.